\documentclass[twocolumn,showpacs,preprintnumbers,floatfix,amsmath,amssymb]{revtex4}
\usepackage{epsfig,psfrag}
\usepackage{dcolumn}     
\usepackage{bm}          
\begin{document}


\date{\today}

\title{Exact results for one-dimensional disordered bosons 
with strong repulsion}

\author{A.~De~Martino,$^1$ M.~Thorwart,$^1$ R.~Egger,$^1$ and R.~Graham$^2$}

\affiliation{$^1$Institut f\"ur Theoretische Physik, 
Heinrich-Heine-Universit\"at, 
D-40225 D\"usseldorf, Germany 
\\ $^2$Fachbereich Physik, Universit\"at Duisburg-Essen, D-45117 Essen, Germany
}

\begin{abstract}
We study one-dimensional disordered bosons with strong repulsive interactions.
A Bose-Fermi mapping expresses this
problem in terms of non-interacting Anderson-localized fermions, whereby
known results for the
distribution function of the local density of states, 
the spectral statistics, and density-density correlations
can be transferred to this new domain of applicability.  
We show that disorder destroys bosonic quasi-long-range order 
by calculating the momentum distribution,
and comment on the experimental observability of these predictions 
in ultracold atomic gases.
\end{abstract}


\maketitle

The properties of interacting bosons  have recently attracted
considerable attention due to the unprecedented control
and tunability achieved in ultracold atomic gases. 
For instance, by using optical lattices, the predicted quantum phase transition
from a superfluid state to a Mott insulator  
\cite{fisher,jaksch} has been experimentally observed \cite{greiner}.  
Current interest is also directed towards disordered systems,
where disorder can be generated by using laser speckle patterns \cite{speckle},
additional incommensurate optical lattice potentials  
\cite{lewenstein,dis1}, or via atom-surface interactions
in micro-chip confined atomic gases \cite{schmied}.
It thus appears feasible to experimentally study
dirty bosons under controlled conditions, in contrast to earlier
realizations using granular superconductors
or Helium-4 in porous media. 
Unfortunately, the theory of disordered interacting bosons is 
difficult, and  no exact solutions are known
apart from numerical or approximate results 
\cite{fisher,ma,rts,singh,dis2,dis2a}, 
even in the one-dimensional (1D) limit \cite{giamarchi,dis3}.  

In this paper, we show that for
strong repulsive interactions, the dirty boson problem in 1D is
exactly solvable via a Bose-Fermi mapping discussed before in 
the clean limit of a Tonks-Girardeau gas 
\cite{tonks,girardeau,lieb,lenard,tracey,haldane}.
The mapping establishes a connection to non-interacting 
disordered fermions,
allowing to directly apply many results previously obtained on Anderson
localization in 1D, and greatly simplifying the
computation of other quantities like the momentum distribution. 
Our predictions can be checked using state-of-the-art
experiments.  Detailed conditions for the 1D regime have been specified in 
 Refs.~\cite{petrov,duniko}, and the 1D
 Tonks-Girardeau regime has recently been achieved \cite{paredes,weiss},
see also Refs.~\cite{axel,moritz}.

The Bose-Fermi
 mapping can be established most directly by starting from a lattice
description of hard-core bosons, the Bose-Hubbard model  \cite{fisher},
which applies immediately to optical-lattice experiments upon expanding
the Bose field operator in the Wannier state basis \cite{jaksch}. 
Considering spinless bosons on a 1D lattice with spacing $a$, with
Bose annihilation operator $b_l$ at site $l$, the Hamiltonian is
\begin{equation}\label{ham}
H=\sum_{l} \left(-J_l \left[b^\dagger_{l+1} b^{}_l+ b^\dagger_l 
b_{l+1}^{}\right] + \epsilon_l  n_l + \frac{ U}{2}  n_l (n_l-1)\right) ,
\end{equation}
where $n_l=b^\dagger_l b_l^{}$.  Here $\epsilon_l= h_l + b l^2$ 
includes a random on-site energy $h_l$ and an 
axially confining harmonic potential,
and $J_l$ is a random hopping amplitude between neighboring sites.
In optical lattices, hopping disorder is suppressed against 
on-site disorder \cite{dis1}, and we thus take
$J_l\equiv J$, but $h_l$ 
distributed according to a Gaussian ensemble \cite{booklifshits} with 
\begin{equation}\label{lattgauss}
\overline{h_l}=0, \quad \overline{ h_l h_{l'} } =   \Delta \delta_{ll'},
\end{equation}
where the overbar denotes the disorder average and $\Delta$ the
disorder strength. Using the spatial 
diffusion constant $D_s$, for a given disorder mechanism,
$\Delta=\hbar^2 v_F^3/(aD_s)$ 
can be expressed in terms of microscopic parameters; $v_F$ is defined
after Eq.~(\ref{fermham}) below.
Detailed theoretical estimates for $D_s$ (and hence $\Delta$) are available for 
laser speckle fields \cite{speckle} and quasiperiodic optical lattices
\cite{lewenstein,dis1}, where also the Gaussian distribution of the disorder 
field $h_l$ is justified. 
We show below that the neglect of disorder in the $J_l$ is no
fundamental restriction.

In the hard-core boson limit, $U\to\infty$, only the occupation 
numbers  $n_l=0$ or 1 are allowed,
and then Eq.~(\ref{ham}) can be mapped to a  non-interacting
lattice fermion model by means of a 
Jordan-Wigner transformation,
$b_l= e^{i\pi\sum_{j<l} c^\dagger_j c^{}_j } c_l ,$
where the $c_l$ denote lattice fermion operators. This 
transformation results in the fermionic Hamiltonian
\begin{equation}                \label{hamF}
H=  \sum_l \left(-J^{}_l \left[c^\dagger_{l+1} c^{}_l + 
c^\dagger_l c^{}_{l+1} \right] + 
 \epsilon^{}_l  c^\dagger_l c^{}_l \right).
\end{equation}
It provides a one-to-one mapping, preserving the Hilbert space structure
of the bosonic problem, with the $N$-particle
bosonic wavefunction expressed in terms of the fermionic
 one as \cite{girardeau}
\begin{equation}\label{mapp}
\Phi_{\nu}^B (l_1, \dots, l_N) =  
 \left| \Phi^F_{\nu} (l_1,\dots,l_N) \right|. 
\end{equation}
The energy level $E_\nu$ for
an $N$-boson eigenstate $\Phi^{B}_\nu$ can thereby be
computed in terms of the 
non-interacting fermionic Hamiltonian (\ref{hamF}). In particular, with the 
single-particle energy $\epsilon^{(j)}_i$ for the $j$th fermion
residing in a single-particle solution $\Psi_i$ to Eq.~(\ref{hamF}), 
and taking into account the exclusion principle,
 $E_\nu=\sum_{j=1}^N \epsilon^{(j)}_i$. 
The many-body fermionic wavefunction 
$\Phi^F_{\nu} (l_1,\dots,l_N)$, and hence also the bosonic one (up to a
sign), is then a Slater determinant,
${\rm det}\left[\Psi_i(l_j)\right]/\sqrt{N!}$.
Since the modulus square does not change under the mapping
(\ref{mapp}), all bosonic quantities given solely
 in terms of $|\Phi^{B}_\nu|^2$ coincide with the fermionic ones.
This includes all correlation functions of the 
particle density and the local density of states (LDoS), 
\[
\rho(\epsilon,l) = \sum_{\nu}\sum_{l_2,\dots,l_N} 
\delta(\epsilon-E_\nu)|\Phi^{B}_\nu(l,l_2,\dots,l_N)|^2 .
\]
The density of states (DoS) per site 
in a lattice with $L$ sites is then
$\rho(\epsilon) = \sum_{l=1}^L \rho(\epsilon,l)/L$, and
also remains invariant.
The same reasoning applies to the continuum 
limit studied later. In particular, the compressibility 
$\kappa$, and thus also the sound velocity, 
is simply 
\begin{equation}\label{compress}
\kappa^{-1}= \frac{\pi^2}{m} \left(\frac{N}{La}\right)^3,
\end{equation}
where $m$ is the atomic mass.
We set $\hbar=1$ and temperature to zero from now on.   

\begin{figure}
\scalebox{0.33}{
\includegraphics{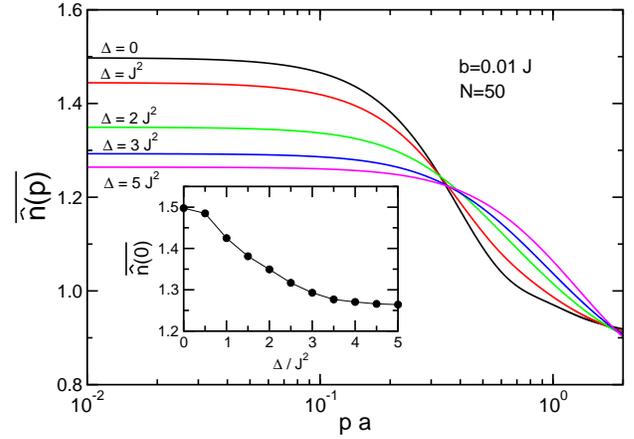}
}
\caption{\label{fig1}  (Color online) Momentum 
distribution (\ref{momentum}) for several $\Delta$
and $N=50$ rubidium atoms in a harmonic axial trap.  
For the disorder average, at most 300 disorder realizations were 
sufficient for convergence, and
$L$ was chosen large enough to ensure $L$-independence.
 Note the linear-logarithmic scale.
Inset: Zero-momentum peak as a function
of disorder strength.
}
\end{figure}

The equality of fermionic and bosonic
results does {\sl not}\ apply to the
momentum distribution,
\begin{equation}\label{momentum}
\overline{\hat{n}(p)}= \frac{1}{N} \sum_{ll'} e^{-ip(l-l')a} 
\overline{\langle b^\dagger_l b^{}_{l'}\rangle} .
\end{equation}
Nevertheless, the Bose-Fermi mapping 
allows for a rather simple exact calculation of the disorder-averaged 
boson momentum distribution.
Using the Jordan-Wigner transformation and
Wick's theorem, Eq.~(\ref{momentum}) for a given
disorder realization can be written as a
 T\"oplitz determinant. For $l>l^\prime$, we find
$\langle b^\dagger_l b_{l'}^{} \rangle = 2^{l-l^\prime-1}
 {\rm det} [G^{(l,l')}]$,
where the $(l-l')\times (l-l')$ matrix has the entries
$G^{(l,l')}_{i,j} = \langle c_{l'+i}^\dagger c^{}_{l'+j-1}\rangle
- \delta_{i,j-1}/2$, see also Ref.~\cite{paredes}.
For fixed disorder $\{ h_l \}$ and arbitrary trap potential, we
compute Eq.~(\ref{momentum}) numerically
and subsequently average over different disorder
realizations.
This is a much faster and more reliable procedure than directly studying
interacting dirty bosons \cite{dis2a,dis3}, since we have  
to deal with a single-particle problem only. 
Let us first consider ${}^{87}$Rb atoms in a harmonic
axial trap with $b=0.01 J$, see Eq.~(\ref{ham}).  
The overall energy scale is set by $J$, which can
be tuned in optical lattices over a wide range \cite{paredes}.
We show results for $N=50$ atoms in Figure \ref{fig1}.
Clearly, disorder has a significant effect on the 
momentum distribution. In particular, 
some weight is transferred to large momenta,
and the zero-momentum peak decreases, see inset in Fig.~\ref{fig1}. 
The momentum distribution has already been measured for bosonic atoms in the
clean 1D limit using Bragg spectroscopy \cite{richard}, and 
through imaging of the
atom cloud after sudden removal of the trap potential \cite{paredes}.
Applying these techniques to the disordered case would allow to test
our predictions.  
Note that these changes in the momentum distribution
reflect Anderson localization physics, and should differ
even qualitatively in the small-$U$ limit. 

\begin{figure}
\scalebox{0.33}{
\includegraphics{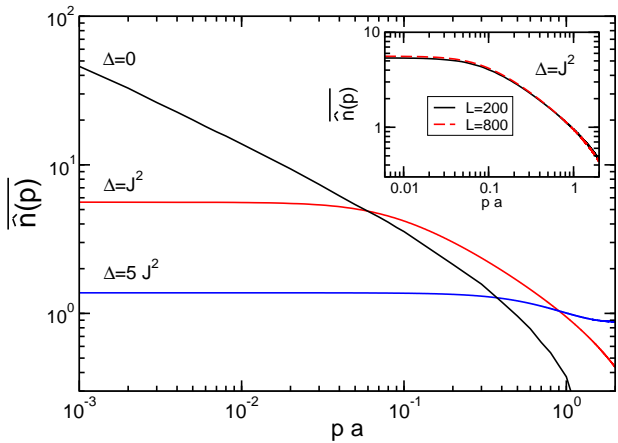} }
\caption{\label{fig2} (Color online) Momentum distribution on a ring 
for various $\Delta$, $N/L=1/2$, with $L=600$. The $\Delta=0$ result
is analytical \cite{lenard,tracey} and shows the $p^{-1/2}$ scaling
at $p\to 0$.  Note the double-logarithmic scale.
The inset shows that for finite $\Delta$ and $L\geq 200$,
finite-size effects are negligible.}
\end{figure}

For a clean homogeneous ($b=0$) system,
the boson momentum distribution 
is well-known to possess a $\hat n(p\to 0)\propto
|p|^{-1/2}$ singularity \cite{lenard,tracey}, corresponding to the one-particle
density matrix $\rho(x,x')\propto |x-x'|^{-1/2}$ for $|x-x'|\to \infty$.  
In the thermodynamic limit, Bose-Einstein
condensation is absent, but there is quasi-long-range order
characterized by the $p^{-1/2}$ law. 
Remarkably, disorder has a fundamental effect on this singular behavior.
The reasoning of Ref.~\cite{klein} allows to prove that 
$\overline{\hat{n}(0)}$  now must remain finite.  
The full momentum distribution $\overline{\hat{n}(p)}$ is
obtained numerically and shown for a ring with periodic boundary conditions
in Fig.~\ref{fig2}.
The complete destruction of quasi-long-range order by disorder is 
clearly visible, and the momentum distribution becomes remarkably
flat for sufficiently strong disorder.

The Bose-Fermi mapping can also be established
via the low-energy theory \cite{giamarchi}, 
which is a perhaps more natural description
for magnetically trapped or micro-chip confined 
atoms, where no underlying lattice is present \cite{schmied}.  
Focussing on circular or hard-wall 
axial trap potentials ($b=0$), corresponding to periodic or open
boundary conditions, the resulting fermionic theory
coincides with the continuum limit ($a\to 0$) of Eq.~(\ref{hamF}).
For a generic incommensurate filling $N/L$, we decompose
the operator $c_l$
into right- and left-moving ($\psi_R,\psi_L$) components with
momenta $k\approx \pm k_F\equiv \pm \pi N/La$ according to 
$c_l \simeq \sqrt{a}\left[ e^{i k_F x} \psi_R(x) + e^{-ik_Fx} \psi_L(x)
\right]$, where $x=la$. 
Correspondingly, the random on-site energies $h_l$ can be
decomposed into a slow part and a term varying on a microscopic scale,  
$h_l \approx \mu (x) + ( \xi (x) e^{-i2 k_F x} + {\rm h.c.})$.
For $h_l=0$, we have a 1D massless Dirac Hamiltonian, 
$\xi(x)$ produces a complex-valued random mass term,
 and $\mu(x)$ a random chemical potential.
Additional disorder in the $J_l$ can be included in 
$\mu(x)$ and $\xi(x)$, and
with bispinor $\psi=(\psi_R,\psi_L)$, the continuum model reads
$H=\int dx \, \psi^\dagger \hat h \psi$ with
\begin{equation} \label{fermham}
\hat h= -i v_F \sigma^z \partial_x +\mu(x) + 
\xi (x) \sigma^+ + \xi^*(x) \sigma^-,
\end{equation}
where $v_F=2aJ\sin(k_F a)$ is the Fermi velocity.
Here the $2\times 2$ matrices $\sigma^{\pm}= (\sigma^x \pm i \sigma^y)/2$ 
are defined in terms of Pauli matrices $\sigma^i$ acting in spinor space.
Equation (\ref{fermham}) is the standard non-interacting 
Hamiltonian used to study Anderson localization in 1D conductors 
\cite{booklifshits,gorkov,altshuler,efetov}.
The forward scattering term $\mu(x)$ can be eliminated by a gauge
transformation and does not affect the quantities of interest below.
(This is  not possible on half-filling, where  $\xi(x)$ is real-valued 
and important differences arise \cite{booklifshits}.)
Equation (\ref{lattgauss}) then implies 
\begin{equation}\label{gaussian}
\overline{\xi(x)} = \overline{\xi^*(x)} = 0 ,  \quad 
\overline{\xi^*(x)\xi(x')} =\frac{v_F}{2\tau } \delta(x-x') ,
\end{equation}
with $\tau=\ell/v_F$ for mean free path $\ell$, 
which also gives the localization length. 
We only discuss weak disorder, $k_F \ell \gg 1$, where 
the bosonic system is in the Bose glass phase \cite{fisher,giamarchi}.
The average DoS is simply
$\bar \rho(\epsilon)=1/(\pi v_F)$ \cite{booklifshits},
and we turn to the LDoS probability distribution. 

The bosonic LDoS (from now on normalized to the average DoS) 
can be expressed in terms of eigenstates $\Psi_{i}(x)$ 
with energy $\epsilon_i$ of the Hamiltonian (\ref{fermham}),
$\rho(\epsilon,x) =\pi v_F  \sum_{i} |\Psi_{i}(x)|^2
\delta(\epsilon-\epsilon_i).$
In a finite and closed sample, these levels are discrete and sharp, and 
it is necessary to regularize the $\delta$-functions. 
A  natural way \cite{altshuler} is to smear out the $\delta$-peaks,
$\rho_f(\epsilon,x)=\int d\epsilon' \rho(\epsilon',x) f(\epsilon-\epsilon')$, 
for instance by using a Lorentzian weight function,
$f_\eta(\epsilon) =  \eta/(\pi[\epsilon^2+\eta^2])$.
Physically, the width $\eta$ is determined by inelastic processes, finite 
sample lifetimes, and escape rates of the trap. 
For an infinite sample, $\rho_f$ then follows 
the inverse Gaussian probability distribution \cite{altshuler} 
\begin{equation} \label{dist1}
W(\rho_f)=\sqrt{\frac{4\eta \tau}{\pi \rho_f^3}}
e^{ -4\eta\tau (\rho_f-1)^2/\rho_f},
\end{equation}
which decreases exponentially both as a function of $\rho_f$ for
$\rho_f \to\infty$, and as a function of $1/\rho_f$ for 
$\rho_f \to 0$.  
The anomalously small probability to find small $\rho_f$ 
implies a Poisson distribution of the energy levels, indicating
the absence of correlations among close-by levels.
This is  obvious in the fermionic picture, 
where energy levels of localized
non-overlapping states cannot repel each other. In the strongly interacting 
bosonic picture, where well-defined single-particle states need not exist, 
this is a much less obvious result.
Let us average $\rho_f(\epsilon,x)$
also over a spatial range $\delta$
determined by the spatial resolution, e.g., the wavelength of a probe laser.
For  $1/k_F\ll \delta \ll \ell$ and $4\eta\tau\ll 1$,
the resulting LDoS $\tilde \rho(\epsilon,x)$
is independent of $\delta$ and
follows the distribution \cite{altshuler}
\begin{equation}\label{dist2}
 \tilde{W}(\tilde\rho)=\frac{\eta \tau}{\pi} \int_4^{\infty}
 dt\ t\  \sin(\pi \eta \tau t) \left( \frac{t+4}{t-4} \right)^{\eta \tau t}
e^{ -\frac{1}{2}\eta \tau \tilde \rho t^2 },
\end{equation}
which is a somewhat narrower distribution than $W(\rho_f)$.
Both Eqs.~(\ref{dist1}) and (\ref{dist2}) remain valid also
in a finite closed sample, as long as the distance to any boundary
is large compared to $\ell$.
The LDoS can be measured using imaging methods or two-photon Bragg spectroscopy 
\cite{stenger,essl}, thus allowing for experimental checks.

We then briefly discuss the bosonic LDoS correlations 
$R(\omega,x)$ at different energies and locations,
\begin{equation} \label{widy}
R(\omega,x-x') = \overline{\tilde{\rho}(\epsilon,x) \tilde{\rho}
(\epsilon+\omega,x')} - 1 ,
\end{equation}
which equal the fermionic ones computed in Ref.~\cite{gorkov}. 
The correlator (\ref{widy}) describes fluctuations in the
spectral statistics related to energy level
repulsion or attraction. It is 
translationally invariant and independent of the
energy $\epsilon$ after the disorder average \cite{efetov}. 
Since $R(\omega,x=x')=0$ 
\cite{gorkov,altshuler,nakhmedov}, we consider
 $k_F|x-x'|\gg 1$  and $\omega \tau \ll 1$, 
where the correlator (\ref{widy}) is finite, with the
limiting values  $R=-1/3$ for small $x$
 and $R= 0$ for $x\to \infty$.  In addition, there is
a deep dip for $\ell \ll x \ll z_0(\omega) = 
2\ell \ln(8/\omega\tau)$,  where
\begin{equation}\label{dip}
R(\omega, \ell \ll x\ll z_0) = -1 + \frac{\pi^{7/2} e^{-x/4\ell}}{16
(x/\ell)^{3/2} } .
\end{equation}
Here $z_0(\omega)$ is the distance two nearly 
degenerate localized states must have to 
generate the energy  splitting $\omega$.
The dip (\ref{dip}) implies that two states with nearly equal energies
occupy with high probability locations far away from
each other. Nevertheless, the wavefunctions of these states must
have an appreciable overlap for short distances, $x\alt \ell$. 
For $x\agt z_0$, the LDoS correlations approach the
uncorrelated limit,
\begin{equation}\label{large}
R(\omega,x\agt z_0) = \frac12 \left[ {\rm erf}\left(\frac{ x-z_0}
{2\sqrt{z_0 \ell}} \right) - 1\right],
\end{equation}
where ${\rm erf}(x)$ denotes the error function.
These features illustrate
that the localized states are centered on many defects,
leading to a complicated quantum interference phenomenon. 
As a consequence, close-lying levels do not obey the
usual Wigner-Dyson spectral correlations found in granular metals
 \cite{efetov},
but instead follow the Poisson statistics of uncorrelated energy levels.  
It would obviously be quite exciting to probe (\ref{widy}) experimentally.
For ultracold bosonic atoms, this is possible using stimulated
two-photon Bragg scattering spectroscopy \cite{stenger}.
Similarly,  one can compute the
(Fourier-transformed) bosonic density-density correlations
$K(x-x',\omega)$, 
\begin{equation}
{\rm Re}\ K(x,\omega+i0) = \pi \int d\epsilon \ n_F(\epsilon)
[1-n_F(\epsilon+\omega)]\
p(x;\epsilon+\omega,\epsilon),
\end{equation}
where $n_F(\epsilon)$ is the Fermi function,
and the fermionic spectral function 
\begin{eqnarray*}
&& p(x-x';\epsilon+\omega,\epsilon) =\\
&& \overline{\sum_{ij}\delta(\epsilon+\omega-\epsilon_i)
\delta(\epsilon-\epsilon_j)\Psi_i(x)\Psi_i^*(x')\Psi^*_j(x')\Psi_j(x)}
\end{eqnarray*}
is a phase-sensitive quantity without a direct bosonic 
image. This nicely illustrates that the Bose-Fermi 
mapping opens otherwise unavailable routes for calculation. 
For $x\agt z_0$ \cite{gorkov},
\begin{equation}
p(x;\epsilon+\omega,\epsilon)=-\frac{
\exp\left[-\frac{(x-z_0)^2}{4z_0\ell}\right] }{2(\pi v_F)^2
\sqrt{\pi z_0/\ell}},
\end{equation}
while for $x\alt z_0$,
to very good approximation \cite{gorkov},
$p(x;\epsilon+\omega,\epsilon) = (\pi v_F)^{-2}(R(\omega,x)+1)$.

To conclude, we have provided exact results for strongly
repulsive dirty bosons in 1D, which can be 
obtained from a Bose-Fermi mapping to non-interacting disordered
fermions.  A Bose-glass phase is thereby mapped to an Anderson-localized
fermionic phase.  A similar mapping is also available for arbitrary 
interaction strength, but involves interacting fermions with
a non-standard contact interaction \cite{shi}.   
For strong (but finite) repulsive bosonic interactions, the weak fermionic
interactions can safely be treated on a perturbative level  \cite{giamarchi}
and cause no substantial differences to our predictions.
Finally, other quantities not discussed here can also be inferred, e.g., the
crossover from short-time diffusive wave-packet expansion 
to localized behavior at long times \cite{nakhmedov}.

We thank A. G\"orlitz and A.O. Gogolin for discussions.
This work was supported by the SFB TR/12 of the DFG.

\end{document}